\documentclass[journal=jctcce,manuscript=article]{achemso} 
\setkeys{acs}{articletitle = true} 

\usepackage[version=3]{mhchem}   
\usepackage[hidelinks]{hyperref}

\author{Maria Ley-Flores}
\affiliation{Pritzker School of Molecular Engineering, University of
Chicago}
\author{Riccardo Alessandri}
\affiliation{Pritzker School of Molecular Engineering, University of
Chicago}
\author{Sean Najmi}
\affiliation{Department of Chemical and Biomolecular Engineering, University of Delaware}
\author{Michele Valsecchi}
\affiliation{Institute for Molecular Science and Engineering, Imperial College London}
\author{George Jackson}
\affiliation{Institute for Molecular Science and Engineering, Imperial College London}
\author{Amparo Galindo}
\affiliation{Institute for Molecular Science and Engineering, Imperial College London}
\author{LaShanda Korley}
\affiliation{Department of Chemical and Biomolecular Engineering, University of Delaware}
\author{Dionisios G. Vlachos}
\affiliation{Department of Chemical and Biomolecular Engineering, University of Delaware}
\author{Juan J. de Pablo}
\affiliation{Pritzker School of Molecular Engineering, University of
Chicago}
\alsoaffiliation{Materials Science Division, Argonne National Lab}

\email{depablo@uchicago.edu}

\title[]{Thermodynamic and Transport Properties of Binary Mixtures of Polyethylene and Higher \emph{n}-Alkanes from Physics-Informed and Machine-Learned Models}

\abbreviations{UA,FF}

\begin{document}

\begin{abstract}
  The thermodynamics and transport properties of polymeric materials are essential for design of reactors and for development of polymer deconstruction processes. Existing property prediction tools such as correlations based on entropy scaling, kinetic gas theory, and free-volume model are inadequate for polymers.
  In this paper, we introduce a data-driven model for polyolefins based on data from molecular dynamics simulations that can accurately predict the transport properties of polyethylenes and their binary mixtures with higher \emph{n}-alkanes across a range of temperatures, pressures, concentrations, and oligomer molecular weights.
  
\end{abstract}


\section{Introduction}
Polyolefins represent the largest fraction of global plastics production\cite{giacovelli2018single,geyer2017production,jambeck2015plastic}. A majority of existing recycling efforts is focused on mechanical recycling processes, which can lead to losses in performance and short usage cycles. Several chemical recycling technologies have been introduced \cite{korley2021toward,epps2021sustainability,hinton2022innovations,zhao2022plastic} to break down polyolefins into smaller molecules, which can then be used for a wide range of potential applications, including as feedstocks for new polymers\cite{coates2020chemical}. Hydrocracking, in particular, is attractive due to the availability of existing technologies that could facilitate industrial-scale adoption.

In polyethylene hydrocracking, precise knowledge of the transport properties of the evolving mixture of polyethylene and cracking products at reactor conditions is essential for developing a fundamental understanding of the kinetics of the deconstruction reactions and the overall process yields. In recent work, Vance \emph{et al}\cite{vance2021single} measured the hydrocracking reactivity of mixtures of $n$-hexacosane ($n$-C$_{26}$) with low-density polyethylene using platinum-tungstated zirconia (Pt-WZr) catalysts. An important conclusion of their measurements was that the addition of $n$-hexacosane affects the conversion of chains having similar or smaller molecular weights.  Thus, it is anticipated that the molecular weight of the cracked products has a pronounced effect on the diffusion of chains into the catalyst pores. To evaluate this hypothesis and to understand the role of transport in polyethylene deconstruction technology, reliable quantitative composition-property relationships are necessary for the estimation of diffusion coefficients at reaction conditions.

A wide range of methodologies and models have been proposed to estimate diffusion coefficients for $n$-alkanes. One approach, suitable for calculating transport properties at low densities, relies on kinetic gas theory, with the Chapman and Enskog theory\cite{taxman1958classical} standing out as a prominent method. An alternative route employs the reduced volume theory developed by Assael and Dymond\cite{assael1990correlation,assael1992correlation,assael1992diffusion}, where they relate the close-packed volume with the number of carbon atoms in alkanes. Other methods based on free-volume modeling and entropy scaling\cite{dehlouz2022entropy} have also been proposed over the last few decades. In addition, several empirical correlations\cite{geet1965prediction,leahy2007molecular,alonso2012determination,von1998diffusion, shieh1969transport, marbach1996self} have been derived from experimental data for $n$-alkanes and their mixtures. However, no single model has been able to reliably relate diffusion coefficients as a function of concentration, chain length, temperature, and pressure for mixtures of $n$-alkanes and polyethylene. Perhaps the main reason for this resides in the constraints of existing experimental tools and equipment, which are unable to measure such properties under extreme conditions, resulting in a scarcity of experimental data for polymer melts and mixtures under reaction conditions.

In the past, molecular dynamics (MD) simulations have been shown to provide an effective tool for the prediction of transport properties under different conditions. By combining Monte Carlo (MC) and MD simulations, Harmandaris \emph{et al}\cite{harmandaris2003crossover} computed values of the self-diffusion coefficient of polyethylenes and determined the crossover from the Rouse regime to the entangled melt regime for linear molecules. Da Silva \emph{et al}\cite{da2020all,da2022all} tested the ability of existing force fields to reproduce experimental properties, including self-diffusion coefficients, of $n$-alkanes and higher $n$-alkanes in the liquid phase. Despite the success of MD simulations in the determination of the liquid-phase properties of hydrocarbons, the computational demands associated with such calculations remain a longstanding challenge. 

Machine learning offers a pragmatic solution to the computational demands of MD simulations. Machine learning has been demonstrated to be a powerful tool for predicting properties of complex hydrocarbon mixtures\cite{dobbelaere2022machine,abdul2018predicting,li2020machine}. Regression models can predict material properties and help streamline processes, saving computational resources and time. Although progress has been made in the application of machine learning methods to polymeric materials\cite{ferguson2017machine,jackson2019recent,chen2021polymer}, the prediction ability of quantitative structure-property relationship (QSPR) models is still limited\cite{audus2017polymer,kumar2019challenges}, mainly due to the absence of sufficiently large and reliable datasets.

To accurately predict physical properties of polyethyene-alkane mixtures over a wide range of compositions (including those resulting from deconstruction processes), we propose the use of statistical models to provide composition-property and structure-property correlations based on values generated using MD simulations at reaction conditions. Statistical models can be trained on a representative subset of data, and then used as predictive models that capture the underlying functions that correlate composition and property.

In this study, we present a model that describes the quantitative relationship between the physically motivated independent variables (i.e., pressure, temperature, polymer concentration, molecular weight of the oligomer) and the diffusion coefficients of binary mixtures of polyethylene and its oligomers. The paper is organized as follows. We begin with a brief description of the molecular model, the simulation methods employed to generate a property data set, and the statistical models used for correlating properties with the composition of the mixtures. Then, we present self-diffusion coefficient results for pure products and binary melts. The extrapolation of the model to the self-diffusion coefficient of single-component $n$-alkanes and polymer melts is compared to available experimental data. Finally, we conclude with remarks concerning the diffusion of oligomers in polymer mixtures, along with the possibilities of the combined molecular simulation–machine learning approach adopted here to study realistic mixtures of industrial interest.

\section{Methods}
\subsection*{Molecular Models}
In this study, we model polyethylenes as monodisperse linear chains of 292 ($n$-C$_{292}$) backbone atoms. Cracked products (polyethylene oligomers) are modeled as linear alkane chains of 20 ($n$-C$_{20}$) up to 100 ($n$-C$_{100}$) backbone atoms. In all cases, MD simulations are performed using an isotropic united-atom model where the hydrogens are implicitly represented.

\subsection*{Simulation Methods}

All MD  simulations were conducted using the package GROMACS 2021.5\cite{van2005gromacs}. One of the main difficulties associated with predictive modeling of polymer mixtures in the liquid phase is the sensitivity of the force field parameters. Accurate prediction of transport properties depends on accurate reproduction of melt densities. Therefore, as a preliminary study, we first test the ability of two of the most accurate force fields\cite{da2020all,da2022all} to reproduce the thermodynamic properties of higher $n$-alkanes: TraPPE-UA\cite{wick2000transferable} and GROMOS 54A7\cite{schuler2001improved, schmid2011definition}. We compare the values with experimental values obtained in literature to the predictions of the SAFT-$\gamma$ Mie equation of state\cite{dufal2014prediction}, which is known to describe accurately the equilibrium thermodynamic properties of alkanes and polyethylene. We then test the self-diffusion coefficient of linear PE and compare it to experimental data.

In conducting all simulations, we modeled the systems of interest using 100 chains, including polymer and oligomer chains. The initial configurations were generated by randomly replacing atoms in a cubic cell 40 nm on a side. After an initial energy minimization at 298 K to remove high-energy configurations from the initial structures, the system was condensed at 523 K and 500 bar in the isothermal-isobaric (NPT) ensemble for 500 ps using the GROMOS 54A7 force field. The last configuration was then used as the starting point for equilibration of individual systems over 10 ns at the corresponding pressure and temperature conditions in the NPT ensemble. Next, the density was fixed to match the prediction of the SAFT-$\gamma$ Mie equation of state, and a 200-ns run was performed at the target temperature in the canonical ensemble (NVT). The diffusion coefficients were computed from the last 100 ns of that production run.

Simulations were performed with a Verlet algorithm\cite{swope1982computer} and a timestep of 1 femtosecond (fs). For condensation, temperature and pressure were controlled using the V-rescale thermostat\cite{bussi2007canonical} and the C-rescale barostat\cite{bernetti2020pressure}. The 10-ns NPT run was performed using the Nosé-Hoover thermostat\cite{nose1984unified,evans1985nose,hoover1985canonical} and the Parrinello-Rahman barostat\cite{parrinello1980crystal,parrinello1981polymorphic}. For NVT simulations, the temperature was controlled using the Nosé-Hoover thermostat. In all cases, the thermostat was fixed with a damping parameter of 100 fs. For all simulations, the cut-off distance for Lennard-Jones interactions was set to 1.4 nm.

The diffusion coefficient of polyethylene in the melt was calculated through the Einstein-Smoluchowski method\cite{islam2004einstein}. First, from the trajectory analysis, the mean square displacement was split into 3 blocks. The diffusion constant was obtained by least squares fitting the mean square displacement in each block to a straight line with the equation: $\langle\Delta r^2 \rangle=6Dt+constant$. The final values for the diffusion coefficients were obtained by averaging the diffusion values obtained for each block. The data for each system were collected for temperatures ranging from 423 to 673 K, pressure between 1 and 50 bar, oligomer molecular weights between 282 g/mol and 1402 g/mol, and polyethylene chains of 4090 g/mol. A total of 530 points were collected for different combinations of temperature, pressure, oligomeric molecular weight, and polymer concentration.

\subsection*{Neural Network}

A feedforward neural network model was constructed using the Sequential API from TensorFlow v2.1.0 \cite{abadi2016tensorflow}, implemented using Keras\cite{chollet2015see}. The model architecture comprised three dense layers with 20 units each, and leaky RELU activation functions were applied to the hidden layers. The number of nodes in the hidden layers was selected through grid search optimization. The input layer had four neurons corresponding to the four input features. The output layer consisted of a single neuron for regression. The model was compiled using the Adam optimizer with a fixed learning rate of 0.001 and mean squared error loss function.

The normalized dataset was split into training and testing sets using a train-test split ratio of 80:20. The model was trained on the training set for 100 epochs with a batch size of 32. A validation split of 0.2 was used during training to monitor the model's performance and prevent overfitting. The number of nodes in the hidden layers was selected through grid search optimization.

The trained model was evaluated using the coefficient of determination (R2 score) on both the training and testing sets. The R2 score was computed to assess the goodness of fit between the predicted and actual diffusion coefficients. Predictions were made on both the training and test sets using the trained model. The predicted values were inverse-transformed to obtain the diffusion coefficients in the original scale for comparison with the actual values.

To provide insights into the model's predictions and understand the impact of input features on the predicted diffusion coefficients of polymer mixtures, SHAP (SHapley Additive exPlanations)\cite{lundberg2017unified} analysis was employed. SHAP is a powerful tool for model interpretation that assigns each feature an importance score indicating its contribution to the model's output. The explainer was instantiated with the trained neural network model and the training dataset. SHAP values were then computed for the testing dataset to explain the model's predictions.


\section{Results and discussion}
\subsection*{Properties of the Single-Component Melt}

Previous work\cite{da2020all,da2022all} has shown that united-atom force fields are better than all-atom force fields for reproducing the thermodynamic and transport properties of hydrocarbons. In particular, TraPPE-UA was found to be superior in reproducing density and GROMOS96 was superior in reproducing self-diffusion coefficients. Importantly, we note here that equations of state (EoS) based on Statistical Associating Fluid Theory (SAFT)\cite{chapman1989saft, chapman1990new} such as the SAFT-$\gamma$ Mie EoS have been shown to be more accurate than available force fields for prediction of the equilibrium thermodynamic properties of $n$-alkanes\cite{shaahmadi2021improving}.

\begin{figure}[h]
    \centering
    \includegraphics[width=1.0\textwidth]{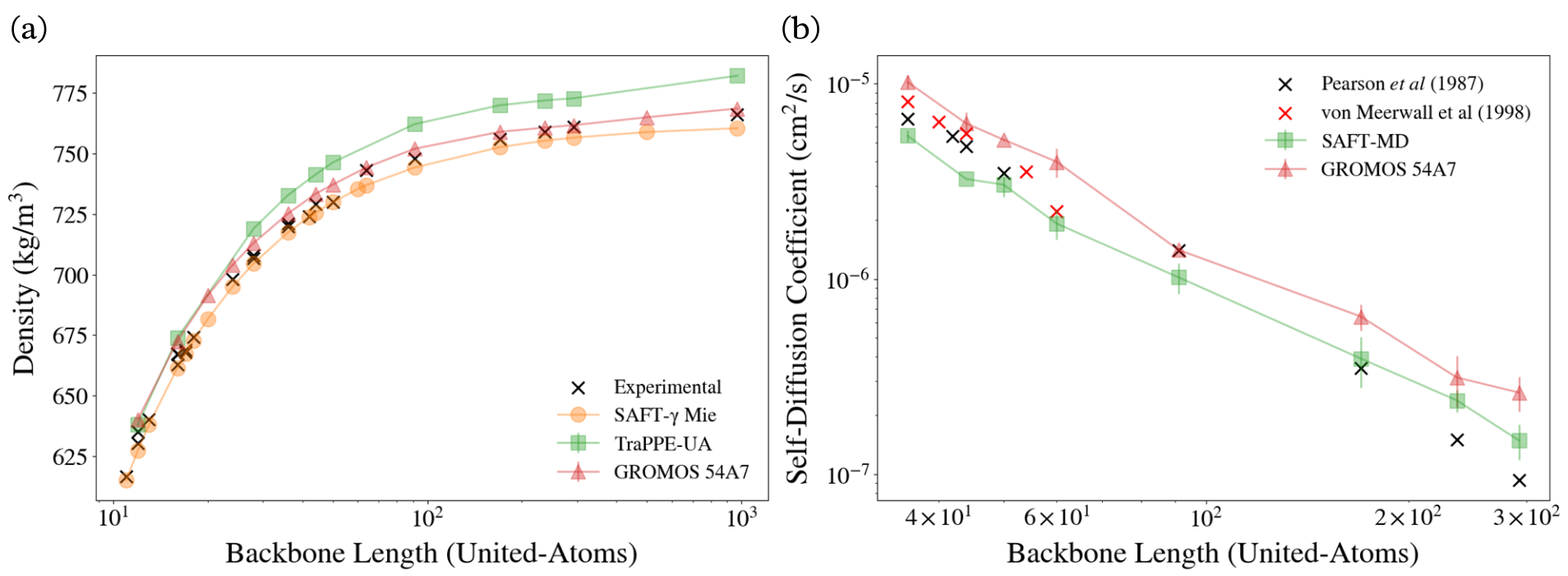}
    \caption{(a) Performance of SAFT-$\gamma$ Mie EoS, TraPPE-UA and GROMOS 54A7 force fields for reproducing experimental density values\cite{pearson1987viscosity, doolittle1951preparation, schuler2001improved, caudwell2004viscosity, tanaka1991viscosity, dymond1980transport} of linear polyethylene at 448.15 K and 1 bar. (b) Self-Diffusion coefficient of linear PE calculated by the SAFT-GROMOS combination and the GROMOS 54A7 force field compared to experimental values \cite{pearson1987viscosity, von1998diffusion}.}
    \label{fig:PE_Density}
\end{figure}

To test the transferability of the most successful forcefields from previous studies to reproduce properties for longer chains (low molecular weight PE), we tested the ability of the GROMOS 54A7 and TraPPE-UA to reproduce the density of polyethylene and compared it with available experimental data \cite{von1998diffusion,pearson1987viscosity} and the predictions of the SAFT-$\gamma$ Mie EoS. The results are shown in Figure \ref{fig:PE_Density}. In general, both the GROMOS 54A7 force field and the SAFT-$\gamma$ Mie EoS are superior to the TraPPE-UA force field in reproducing density for the molecular weights considered here. Furthermore, differences between experimental values and the predictions of TraPPE-UA become more important as the molecular weight increases. For higher $n$-alkanes of up to around 50 carbons, our results indicate that the SAFT-$\gamma$ Mie EoS is superior to the GROMOS 54A7 force field in reproducing liquid density. However, for oligomers longer than 50 methylene units, the GROMOS 54A7 force field performs better than the equation of state, indicating that in this case, the GROMOS 54A7 force field is more suitable for the study of thermodynamic properties of polyethylene. Given that our study is concerned with mixtures that include high concentrations of $n$-alkanes, the prediction of the SAFT-$\gamma$ Mie EoS provides the most reliable values for the density of the mixtures.

Figure \ref{fig:PE_Density}(b) shows the agreement between the self-diffusion coefficient computed using the GROMOS 54A7 force field starting at the density predicted by the SAFT-$\gamma$ Mie EoS and the diffusion predictions starting for the density obtained using the GROMOS 54A7 force field. 
We find that for chains consisting of up to 50 methylene units, the GROMOS force field slightly overestimates the diffusion coefficient, whereas simulations starting from the corrected density using the SAFT-$\gamma$ Mie EoS tend to underestimate it. However, for chains longer than 60 methylene units, simulations at the density predicted by the SAFT-$\gamma$ Mie EoS exhibit better agreement with the experimental self-diffusion values available in the literature.

\subsection*{Machine Learning Model}

\begin{figure}[h]
    \centering
    \includegraphics[width=0.9\textwidth]{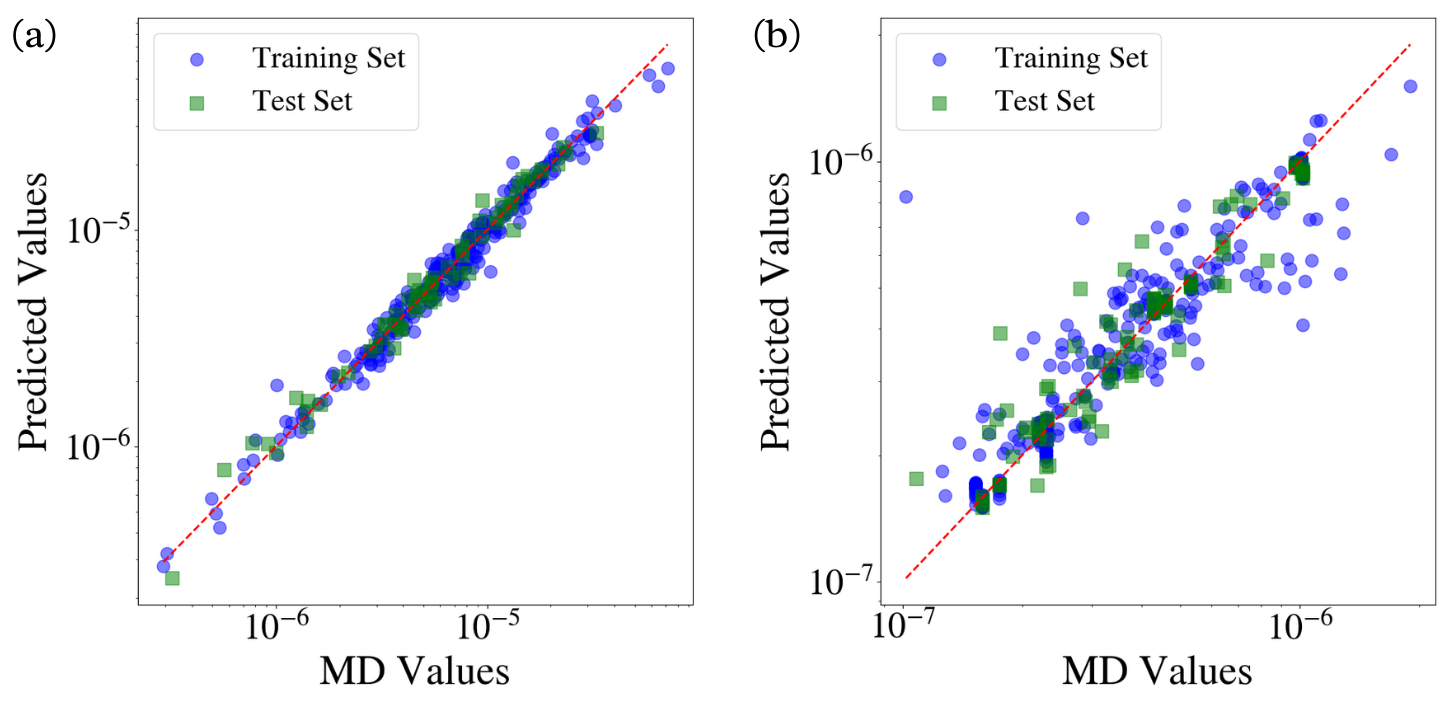}
    \caption{(a) Predictions for the diffusion of the oligomer in the binary mixture. (b) Predictions for the diffusion of the polymer in the binary mixture.}
    \label{fig:mixtures_model}
\end{figure}

Figure \ref{fig:mixtures_model} shows the agreement between simulated and predicted values for the diffusion coefficient of mixtures of $n$-C$_{292}$ and oligomers ranging from $n$-C$_{20}$ and $n$-C$_{100}$ at temperatures ranging from 423 to 673 K and pressures from 1 to 50 bar. The model's predictions for the diffusion coefficient of the oligomer in the mixtures are excellent, with coefficients of determination for the test and training sets of 0.9800 and 0.9795, respectively. The model's predictions for the polymer diffusion are in good agreement with the data, with coefficients of determination for the test and training sets of 0.8787 and 0.8770, respectively. The model compensates for the estimation errors that arise from our high-throughput simulation methodology.

\begin{figure}[h]
    \centering
    \includegraphics[width=0.9\textwidth]{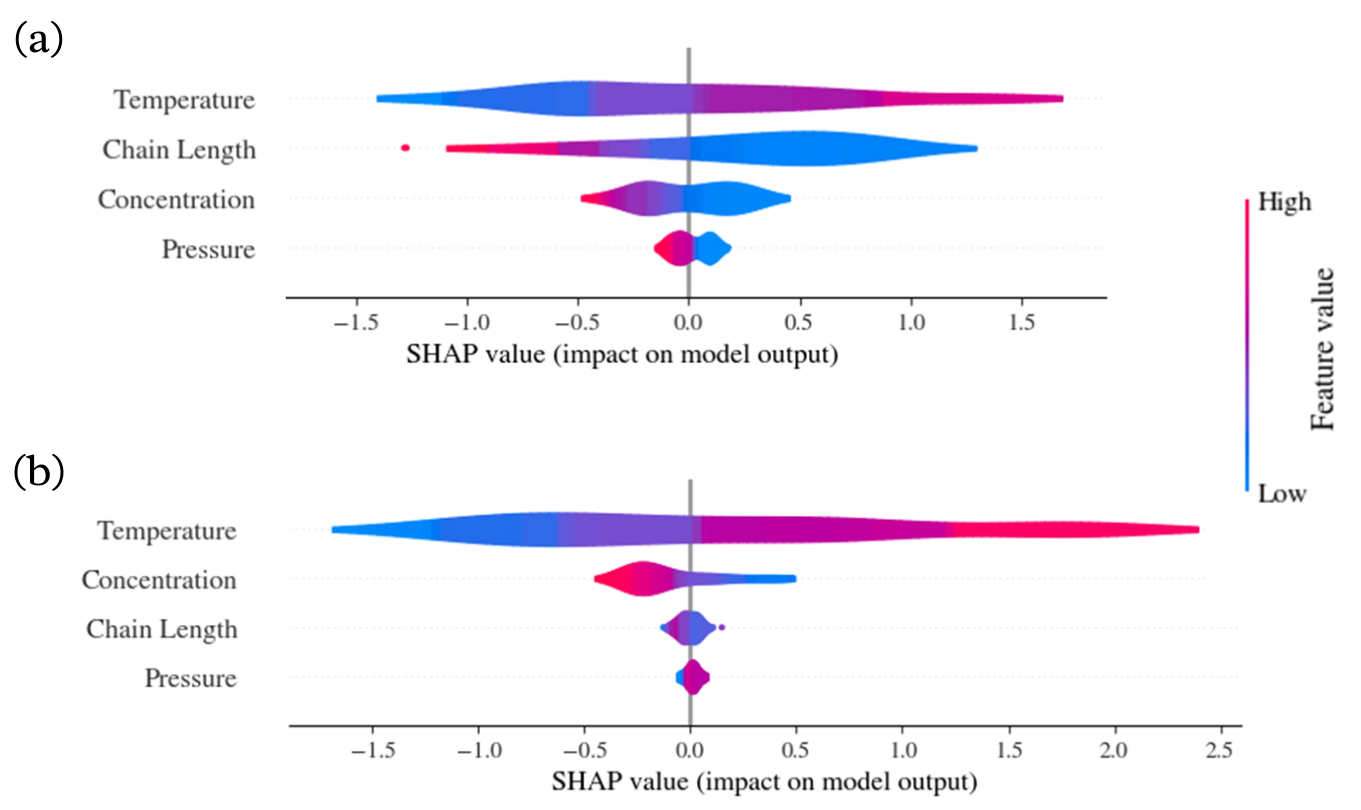}
    \caption{SHAP summary plot for the prediction of the diffusion of (a) oligomer in the mixture and (b) polymer in the mixture.}
    \label{fig:shap}
\end{figure}

Figure \ref{fig:shap} shows a SHAP\cite{lundberg2017unified} summary plot of the model. The width of the violin plot at each feature's position represents the distribution of SHAP values for the feature across the dataset. Wider sections indicate higher variability in SHAP values, suggesting that the feature has a more significant impact on predictions across different instances. The position of the line on the x-axis indicated the impact of the feature on the model's output. Features on the right side contribute positively to increasing the model's prediction, while features on the left side contribute negatively. In this model, temperature, followed by the chain length and concentration are the most important features, with high-temperature values contributing positively to the prediction of diffusion and low-temperature values contributing negatively. The opposite is true for the chain length and concentration, where longer oligomer chain lengths and higher polymer concentrations contribute negatively to the diffusion, while shorter chain lengths and lower polymer concentrations contribute positively.

\begin{figure}[h]
    \centering
    \includegraphics[width=1.0\textwidth]{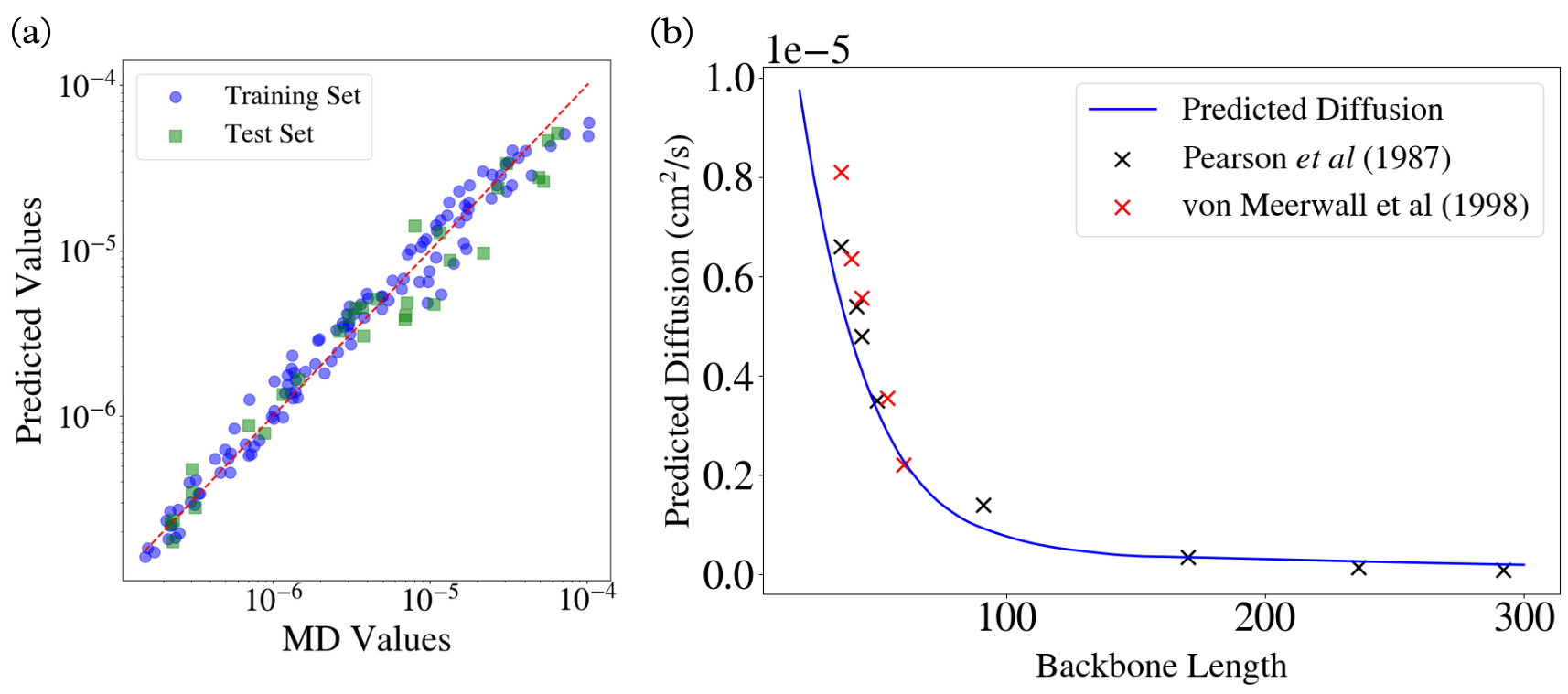}
    \caption{(a) Agreement between the diffusion values of single-component melts computed from MD simulations and the predictions of our model. (b) Predictions for the diffusion of higher $n$-alkanes and polyethylene as a function of the number of carbons in the backbone length compared to experimental values available in the literature\cite{pearson1987viscosity,von1998diffusion}.}
    \label{fig:melt_model}
\end{figure}

Figure \ref{fig:melt_model} compares the diffusion coefficient of the single components extrapolated from the model to single-component melts. The predictions of the model are excellent, with coefficients of determination for the test and training sets of 0.9682 and 0.9691, respectively. Experimental values are compared to the predictions of the model with a coefficient of determination of 0.7473.

\section{Conclusions}

In this study, we successfully integrated a high-throughput molecular dynamics simulation pipeline and a feedforward neural network to establish quantitative relationships for the diffusion coefficients of binary mixtures comprising linear polyethylene and its oligomers. The diffusion coefficients of the individual components in the polymer mixtures show a behavior whose trends with temperature, pressure, and polymer concentration are nontrivial to correlate, serving to complicate the prediction of diffusion coefficients over a wide range of conditions. Nevertheless, our proposed neural network exhibits satisfactory predictive performance, and is well aligned with experimental values reported in the literature. 

A SHAP analysis provided a quantitative measure of the relative feature importance driving the model's prediction. Notably, temperature emerged as the predominant determinant of diffusion for both polymer and oligomer components, with higher temperatures correlating with increased diffusion rates within the mixtures. Conversely, the molecular weight of oligomers exhibited a significant, but opposite, impact on diffusion coefficients, with longer chains corresponding to slower diffusion rates. Interestingly, the influence of oligomer molecular weight on polymer diffusion proved comparatively minor. Additionally, our analysis revealed that polymer concentration plays a more critical role in oligomer diffusion within the mixture, whereas pressure exerted a dampening effect on both oligomer and polymer diffusion coefficients.

Overall, our model's predictions underscore the nuanced interplay between concentration, oligomer molecular weight, and pressure in shaping the diffusion dynamics of polyethylene and its oligomers within mixtures. Specifically, while factors such as concentration and pressure significantly influence oligomer diffusion, the diffusion of polyethylene is primarily governed by temperature, with a relatively minor influence from oligomer molecular weight. These findings not only contribute to a deeper understanding of diffusion behavior in polymer mixtures but also offer practical implications for the design of chemical recycling processes, where precise knowledge of transport properties is crucial for optimizing product yield. Thus, the constructed model and its associated insights stand as valuable tools for guiding future design and optimization of polyolefin deconstruction processes.

\begin{acknowledgement}

This work was supported as part of the Center for Plastics Innovation, an Energy Frontier Research Center funded by the U.S. Department of Energy, Office of Science, Basic Energy Sciences at the University of Chicago under Grant Number DE-SC0021166. 

\end{acknowledgement}

\bibliography{MyBib}

\end{document}